\documentstyle[sprocl]{article}

\input{psfig}

\def\be{\begin{equation}}
\def\ee{\end{equation}}
\def\bea{\begin{eqnarray}}
\def\eea{\end{eqnarray}}

\begin{document}

\title{Testing the Maldacena conjecture with SDLCQ}

\author{Uwe Trittmann}

\address{Department of Physics, Ohio State University\\ 
174 W 18th Ave, Columbus, OH 43210, USA}

\maketitle\abstracts{We report on a test of the 
Maldacena conjecture \cite{plht}. This string/field theory correspondence has 
interesting applications. When combined with Rehren's theorem, it 
has implications
for issues concerning space-time structure and Lorentz symmetry.
Our results indicate that the conjecture is correct. 
We are within 10-15\% of the expected results, although the numerical 
evidence is not yet decisive.}

\section{Introduction}

The so-called Maldacena conjecture \cite{Maldacena}, namely the assertion
that four-dimension\-al 
${\cal N}$=4 supersymmetric Yang-Mills theory (SYM$_{3+1}$)
can be identified in some limits
with type IIB string theory on an $AdS_5\times S^5$ background, has caused
a lot of excitement in recent years. 
Interesting consequences arise when this conjecture is combined with 
Rehren's theorem, which states that
if SYM$_{3+1}$ is an algebraic quantum field theory, 
it induces a dual algebraic field theory 
on an $AdS_5$ space-time with fixed causal structure.
As was pointed out by Arnsdorf and Smolin \cite{Smolin2001}, 
one of the following must then be true.
Either SYM$_{3+1}$ 
is not an algebraic quantum field theory, since it violates the 
causal structure of four-dimensional Minkowski space,
or string theory on $AdS_5\times S^5$ is not a quantum theory of gravity,
because it is equivalent to a quantum field theory 
with fixed causal structure, or
there is no consistent quantization of string theory on  
$AdS_5\times S^5$ for finite string length and coupling.
These findings seem to have important implications for our view 
on the structure of space-time and Lorentz symmetry.
Since Rehren's work contains a strict proof, whereas Maldacena's 
conjecture is a falsifiable hypothesis, 
it is obvious that there is a need for a 
rigorous test of the latter.


In order to test the Maldacena conjecture, one would ideally have the 
following requirements fulfilled at the matching point of the field and 
string theory: small curvature to be able to work in the supergravity 
(SUGRA) approximation to string theory and small coupling in order to 
use perturbation theory on the field theory side.
There is, however, no appropriate scenario known, where both requirements 
would be met.
The way out of this dilemma is to use a non-perturbative method, namely 
supersymmetric discretized light-cone quantization (SDLCQ) in low dimensions,
where it is known to work best. Fortunately, a scenario where a string 
theory corresponds to a low-dimensional field theory is available.
A system of D1 branes in type IIB string theory decoupling from gravity
is conjectured to be dual to ${\cal N}=(8,8)$
supersymmetric Yang-Mills theory in 1+1 dimensions \cite{Itzhaki98}.
We will use the correlation function of a gauge invariant operator, 
namely the stress-energy tensor $T^{\mu\nu}$, as an observable that can be 
computed on both sides of the correspondence, and therefore can be used 
to test the Maldacena conjecture.

\section{The Correlator from SUGRA}

To determine the two-point correlation function of the 
stress-energy tensor from
string theory, one uses the supergravity, {\em i.e.~}small curvature 
approximation.
We do not have room here to go into the details of the 
computation 
and refer the reader to the literature 
\cite{ItzhakiHashimoto}. It suffices to state that 
the leading non-analytic term in the flux factor yields the correlator
\be
\langle {\cal O}(r){\cal O}(0)\rangle=\frac{N_c^{3/2}}{g r^5}.
\ee
As a consistency check we remark that in two-dimensional 
${\cal N}=(8,8)$ SYM one 
has conformal fixed points at the ultraviolet and infrared with central 
charges $N_c^2$ and $N_c$, respectively. One expects to deviate 
from the conformal $1/r^4$ behavior of the correlator at distances 
$r={1}/{g\sqrt{N_c}}$ and $r={\sqrt{N_c}}/{g}$.
This yields the phase diagram depicted below. We shall be 
interested in reproducing the cross-over from the small to the intermediate
distance 
regime, where the correlator changes its behavior from $1/r^4$ to $1/r^5$.
The agenda is then to detect a $1/r$ slope when evaluating the correlator 
at increasing distances on the field theory side.

\vspace{0.15cm} 
\centerline{
\unitlength0.8cm
\begin{picture}(15,2)\thicklines
\put(0.2,1){\vector(1,0){14.8}}
\put(0.2,0.8){\line(0,1){0.4}}
\put(5,0.8){\line(0,1){0.4}}
\put(10,0.8){\line(0,1){0.4}}
\put(0,0){$0$}
\put(4.5,0){$\frac{1}{g\sqrt{N_c}}$}
\put(9.5,0){$\frac{\sqrt{N_c}}{g}$}
\put(14.8,0.2){$r$}
\put(2,0.3){UV}
\put(6.5,0.3){SUGRA}
\put(12,0.3){IR}
\put(1.5,1.5){$N_c^2/r^4$}
\put(6.0,1.5){$N_c^{3/2}/(g r^5)$}
\put(11.5,1.5){$N_c/r^4$}
\end{picture}
}


\section{The correlator from SDLCQ}

Discrete light-cone quantization (DLCQ) \cite{BPP} is known to 
preserve supersymmetry \cite{Sakai95}. This makes it possible to avoid  
the inherent severe renormalization problems in this Hamiltonian 
approach to quantum field theory, given enough supersymmetry. 
The method goes under the name of supersymmetric DLCQ, or SDLCQ. 

To reproduce SUGRA scaling relation, and to calculate the 
cross-over behavior  of the correlator at intermediate 
distances 
using SDLCQ, we have to compute the correlator
\be
F(x^-,x^+)=\langle{\cal O}(x^-,x^+){\cal O}(0,0)\rangle,
\ee
where we introduced the light-cone 
coordinates $x^\pm\equiv\frac{1}{\sqrt{2}}(x^0\pm x^1)$.
As an operator 
we consider the gauge invariant (two-body) operator $T^{++}(-K)$, a component 
of the stress-energy tensor.
In DLCQ one fixes the total longitudinal momentum, $P^+={K\pi}/{L}$,
so we Fourier transform, and decompose the result into momentum modes. Finally,
we continue to Euclidean space by taking distance 
$r^2 = 2 x^+ x^-$ to be real. With the harmonic resolution $K$ playing the 
role of a discretization parameter, we are supposed to send 
$K\rightarrow\infty$ to recover the continuum limit.
We obtain the functional form of the correlator
\be
{\cal F}(r)=\left({x^- \over x^+}\right)^2 F(x^-,x^+)=
\left|{L \over \pi} \langle n | T^{++}(-K) |0 \rangle \right|^2
{M_n^4 \over 8 \pi^2 K^3} {\cal K}_4(M_n r),\label{corr}
\ee
with the mass eigenvalues $M_n$ and a modified Bessel function ${\cal K}_4$.
Note that this result is $K$ dependent, 
but involves no other unphysical quantities. In particular, the box length
$L$ will drop out. 
We obtain the correct small $r$ behavior 
\be
{\cal F}(r) 
\quad{\longrightarrow}\quad
{(2 n_b + n_f) \over 4 \pi^2}
\left(1 - {1 \over K}\right)\frac{N_c^2}{r^4}.
\ee
\vspace*{-0.6cm}

\section{Results}

The correlator, Eq.~(\ref{corr}), is determined by numerical calculation of
mass spectrum, $M_n(K)$, of ${\cal N}=(8,8)$ SYM. 
Amongst the problems we face with the numerical approach 
is that due to the large number of particle species in the theory,
the Fock space grows very fast with the harmonic resolution:
$K=2,3,4$ implies a dimension of the Hamiltonian of $256,1632,29056$.
The necessary improvements on numerical treatment include
the use of a C++ code (more efficient data structure), use of 
the discrete flavor symmetry,
and improvements on numerical efficiency (improved Lanzcos algorithm).
Another problem is the occurrence of 
massless unphysical states. The number of partons in these artifacts
is even (odd) for $K$ being even (odd). Since the 
correlator is only sensitive to two particle contribution, the 
curves ${\cal F}(r)$ will show a different behavior for even and odd 
$K$ in the region where the approximation breaks down.
The problem is that the 
unphysical states yield the correct 
$1/r^4$ behavior, but have a wrong $N_c$ dependence, which prohibits
the detection of the regular contribution at large $r$, which is 
down by $1/N_c$.
We can, however, take the different behavior of the even and odd $K$ curves
to establish where the approximation breaks down.
The continuum limit seems sound, because 
the breakdown of the approximation occurs at larger $r$ as $K$ grows.

\begin{figure}[H]
\vspace*{-1cm}
\centerline{
\psfig{file=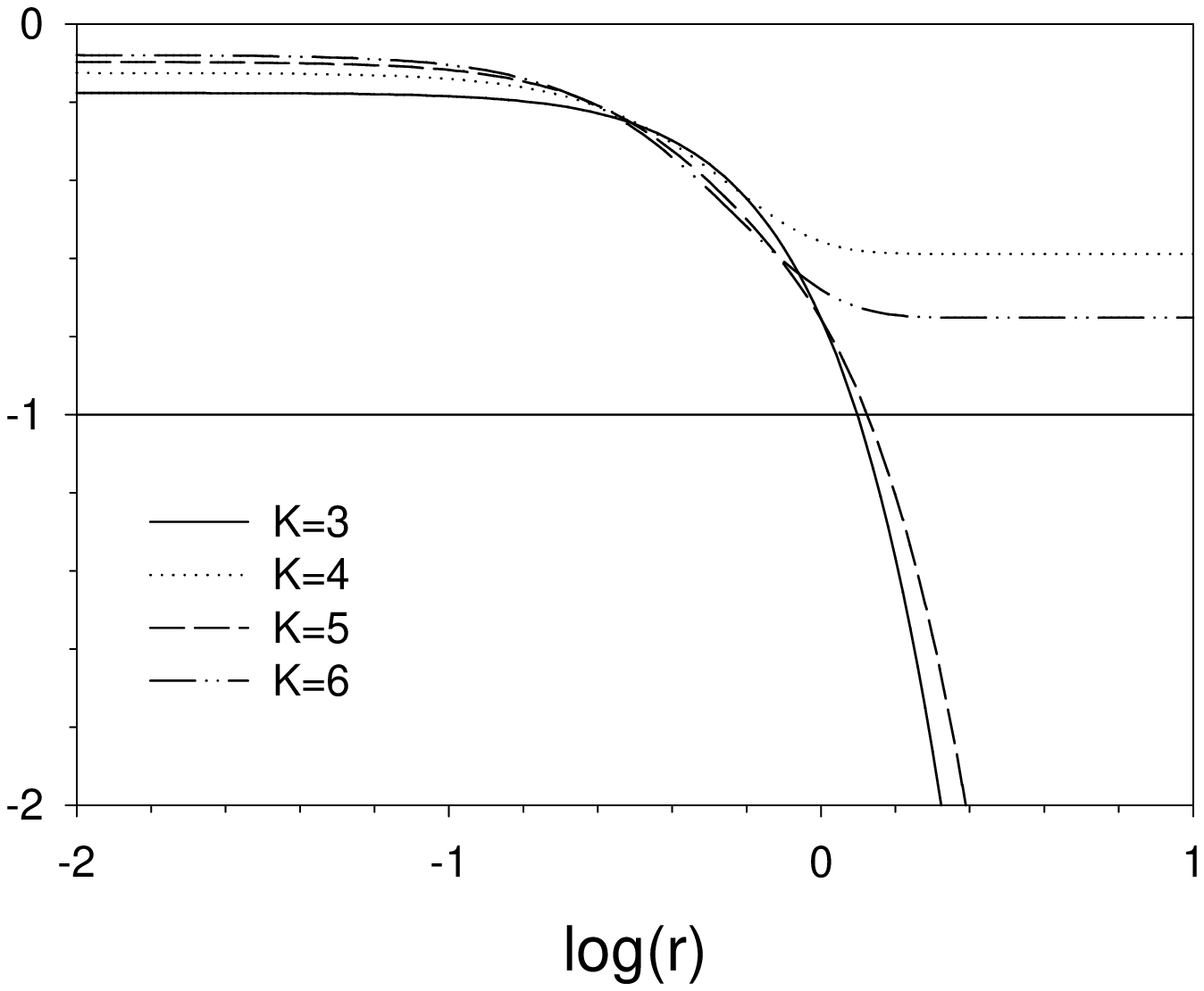,width=10true cm,angle=0}
}
\vspace*{-0.37cm}
\centerline{
\psfig{file=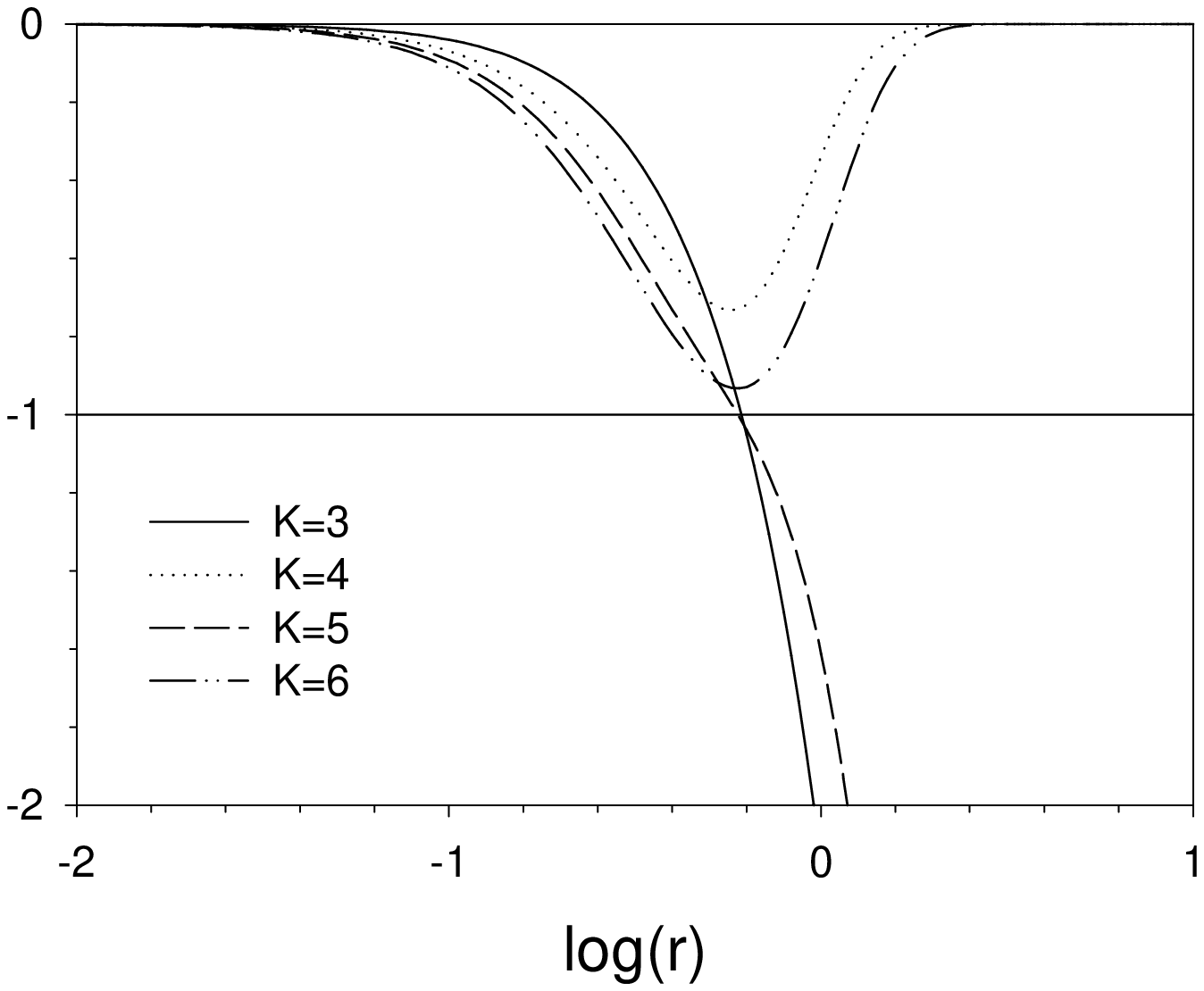,width=10true cm,angle=0}}
\caption{(a) Top: log-log
plot of correlator $\langle T^{++}(x) T^{++}(0) \rangle
\left({x^- \over x^+} \right)^2 {4 \pi^2 r^4 \over N_c^2 (2 n_b +n_f)}$
v.s. $r$ in units $g^2 N_c /\pi$ for $K=3,4, 5$ and
$6$. (b) Bottom: the log-log
derivative with respect to $r$ of the correlation function in (a).
\label{Fig1}
}
\end{figure}

Our expectations are then the following.
The behavior of the correlator ${\cal F}(r)$ changes  
like $1/r^4\rightarrow 1/r^5$ as $r$ increases, so we 
should approach $d{\cal F}/dr=-1$ in the continuum limit.
Hence, we would claim success if the curve $d{\cal F}/dr$ 
flattens at $-1$ before the approximation breaks down.
A look at Figs.~\ref{Fig1}(a) and (b), reveals that these expectations are 
realized. There is a clear tendency of the curves in Fig.~\ref{Fig1}(a)
to develop a negative slope of order unity as $K$ increases. 
To allow for a more quantitative
statement, we plotted the derivative of the curves in Fig.~\ref{Fig1}(b).
Keeping in mind that our approximation breaks down when the odd and even 
$K$ curves cross, we see that the values of the correlator at this point seem
to converge towards unity, as $K$ grows. 

\section{Conclusions and Outlook}

In order to test the Maldacena conjecture, we calculated the 
correlator of the stress-energy tensor on the field theory side
of this string/filed theory correspondence with the non-perturbative SDLCQ  
approach. Our results are within 10-15\% of the results 
expected from Maldacena conjecture. 
The present study includes a factor 100-1000 more states than previously
considered \cite{Hashi}. 
Improvements of code and the numerical method are possible 
and are either on the way, or have been implemented already.
Empirically, we found that
the contributions to the matrix elements come from small number of terms.
An analytic understanding of this fact 
would greatly accelerate calculations
and help improving the test of the conjecture.

As an outlook, we state that for the final goal, namely to test 
the Maldacena conjecture proper 
(${\cal N}=4$ SYM$_{3+1}$ vs.~type IIB string theory 
on AdS$_5\times S_5$), we have to 
apply the SDLCQ approach in larger dimensions. 
This has been partly achieved already in a series of papers of Pinsky,
Hiller, and the author \cite{hpt,hpt01}
with novel numerical improvements by orders of magnitude.

\end{document}